\documentclass[sigconf]{acmart}
\usepackage{multirow}
\AtBeginDocument{%
  \providecommand\BibTeX{{%
    \normalfont B\kern-0.5em{\scshape i\kern-0.25em b}\kern-0.8em\TeX}}}

\copyrightyear{2023}
\acmYear{2023}
\setcopyright{acmlicensed}
\acmConference[MM '23] {Proceedings of the 31st ACM International Conference on Multimedia}{October 29--November 3, 2023}{Ottawa, ON, Canada.}
\acmBooktitle{Proceedings of the 31st ACM International Conference on Multimedia (MM '23), October 29--November 3, 2023, Ottawa, ON, Canada}
\acmPrice{15.00}
\acmISBN{979-8-4007-0108-5/23/10}
\acmDOI{10.1145/3581783.3612420}
\begin{document}

\title{Interactive Interior Design Recommendation via Coarse-to-fine Multimodal Reinforcement Learning}
\author{He Zhang}
\affiliation{%
  \institution{Thrust of Artificial Intelligence, The Hong Kong University of Science and Technology (Guangzhou)}
  \city{}
  \country{}
}
\email{hzhang757@connect.hkust-gz.edu.cn}

\author{Ying Sun}
\authornote{Corresponding authors.}
\affiliation{%
  \institution{Thrust of Artificial Intelligence, The Hong Kong University of Science and Technology (Guangzhou)}
  \city{}
  \country{}
  }
\email{yings@hkust-gz.edu.cn}

\author{Weiyu Guo}
\affiliation{%
  \institution{Thrust of Artificial Intelligence, The Hong Kong University of Science and Technology (Guangzhou)}
  \city{}
  \country{}
  }
\email{guoweiyu96@gmail.com}

\author{Yafei Liu}
\author{Haonan Lu}
\affiliation{%
  \institution{OPPO Research Institute}
  \state{}
  \country{}
  }
\email{liuyafei@oppo.com}
\email{luhaonan@oppo.com}


\author{Xiaodong Lin}
\affiliation{%
  \institution{Rutgers University}
  \state{}
  \country{}}
\email{lin@business.rutgers.edu}

\author{Hui Xiong}
\authornotemark[1]
\affiliation{%
  \institution{Thrust of Artificial Intelligence, The Hong Kong University of Science and Technology (Guangzhou)}
  \institution{Guangzhou HKUST Fok Ying Tung Research Institute}
  \city{}
  \country{}
  }
\email{xionghui@ust.hk}

\renewcommand{\shortauthors}{He Zhang, et al.}
\newcommand{\weiyu}[1]{{\color{blue}[weiyu: #1]}}
\begin{abstract}
Personalized interior decoration design often incurs high labor costs. Recent efforts in developing intelligent interior design systems have focused on generating textual requirement-based decoration designs while neglecting the problem of how to mine homeowner’s hidden preferences and choose the proper initial design. To fill this gap, we propose an Interactive Interior Design Recommendation System (IIDRS) based on reinforcement learning (RL). IIDRS aims to find an ideal plan by interacting with the user, who provides feedback on the gap between the recommended plan and their ideal one. To improve decision-making efficiency and effectiveness in large decoration spaces, we propose a Decoration Recommendation Coarse-to-Fine Policy Network (DecorRCFN). Additionally, to enhance generalization in online scenarios, we propose an object-aware feedback generation method that augments model training with diversified and dynamic textual feedback.
Extensive experiments on a real-world dataset demonstrate our method outperforms traditional methods by a large margin in terms of recommendation accuracy. Further user studies demonstrate that our method reaches higher real-world user satisfaction than baseline methods.
\end{abstract}

\ccsdesc[500]{Information systems~Recommender systems}

\keywords{Multimodal Interaction, Interactive Recommendation, Reinforcement Learning, Interior Design}


\maketitle

\section{INTRODUCTION}
As interior design becomes increasingly prevalent in our daily life, designing a decoration plan that satisfies the personalized requirements of the homeowner has become a critical challenge. Effective communication can be hindered by the homeowner's lack of professional design knowledge and a clear idea of the desired decoration effect. Consequently, a significant amount of effort is spent repeatedly redesigning and rendering the decoration plan to collect concrete feedback on the owner's preferences, resulting in extremely high labor and time costs.

In recent years, there has been an emerging trend in developing intelligent indoor design facilitation systems. Along these lines, major research efforts are being devoted to generating decoration effects based on textual descriptions \cite{wang2021sceneformer,wei2023lego,tang2023diffuscene,gu2022vector}. For example, Text2Room \cite{hollein2023text2room} generates 3D indoor scenes according to the input text, Text2Scene \cite{tan2019text2scene} recurrently generates objects and attributes in the current scene according to the input text. These methods can generate high-quality decoration displays when people have determined their specific requirements and can explicitly describe them. However, there is an undeniable problem: users may not be clear about what they want at the beginning. The problem of how to identify a user’s hidden preferences and choose the proper initial design has been largely neglected.

Recommending the interior designs according to the inherent user preferences is a non-trivial task. First, non-expert users typically require multiple rounds of interaction to express their potential desires in terms of their instant likes and dislikes of presented decoration effects \cite{xu2021adapting,lei2020interactive,lei2020estimation}. Learning an effective and sustainable interaction strategy for selecting optimal decoration plans from a large space that inspires and guides users to express their preferences is a challenging task. The interior decoration design interaction process involves users providing language feedback regarding visual decoration effects. Modeling hidden user preferences from complex multimodal semantic interactions \cite{turk2014multimodal,liao2018interpretable,long2016composite, tan2019lxmert} is another challenging task. Third, real-world user feedback exhibits subjectivity, imprecision, uncertainty, and diversity, which can bring difficulty for model training \cite{jawaheer2014modeling,kotkov2018investigating}.

To address these challenges, we propose an Interactive Interior Design Recommendation System (IIDRS).
\footnote{The dataset and source codes are available at \url{https://github.com/ZhHe11/IIDRS}.}
This system aims to find an ideal plan through limited rounds of interactive recommendation to a user who provides textual feedback on the gap between the recommended plan and the ideal one. Specifically, we propose a novel Decoration Recommendation Coarse-to-Fine Policy Network (DecorRCFN). Observing the multimodal interaction history, DecorRCFN employs Reinforcement Learning (RL) \cite{williams1992simple} to learn the interactive recommendation strategy that explores user preferences and fulfills their requirements. In particular, a coarse-to-fine two-policy structure is proposed to raise efficiency and effectiveness of decision-making on the large decoration action space. 
Furthermore, to enhance model generalization in online scenarios that are characterized by uncertain and diversified feedback, we design an object-aware feedback generation method which augments model training with dynamic and diversified textual feedback. Finally, we conduct extensive experiments on a real-world decoration dataset. The experimental results show that DecorRCFN outperforms existing methods by a large margin in terms of recommendation accuracy. Moreover, user studies demonstrate that our method achieves the highest level of user satisfaction in real-world interactions.

\begin{figure}
  \includegraphics[width=0.35\textwidth]{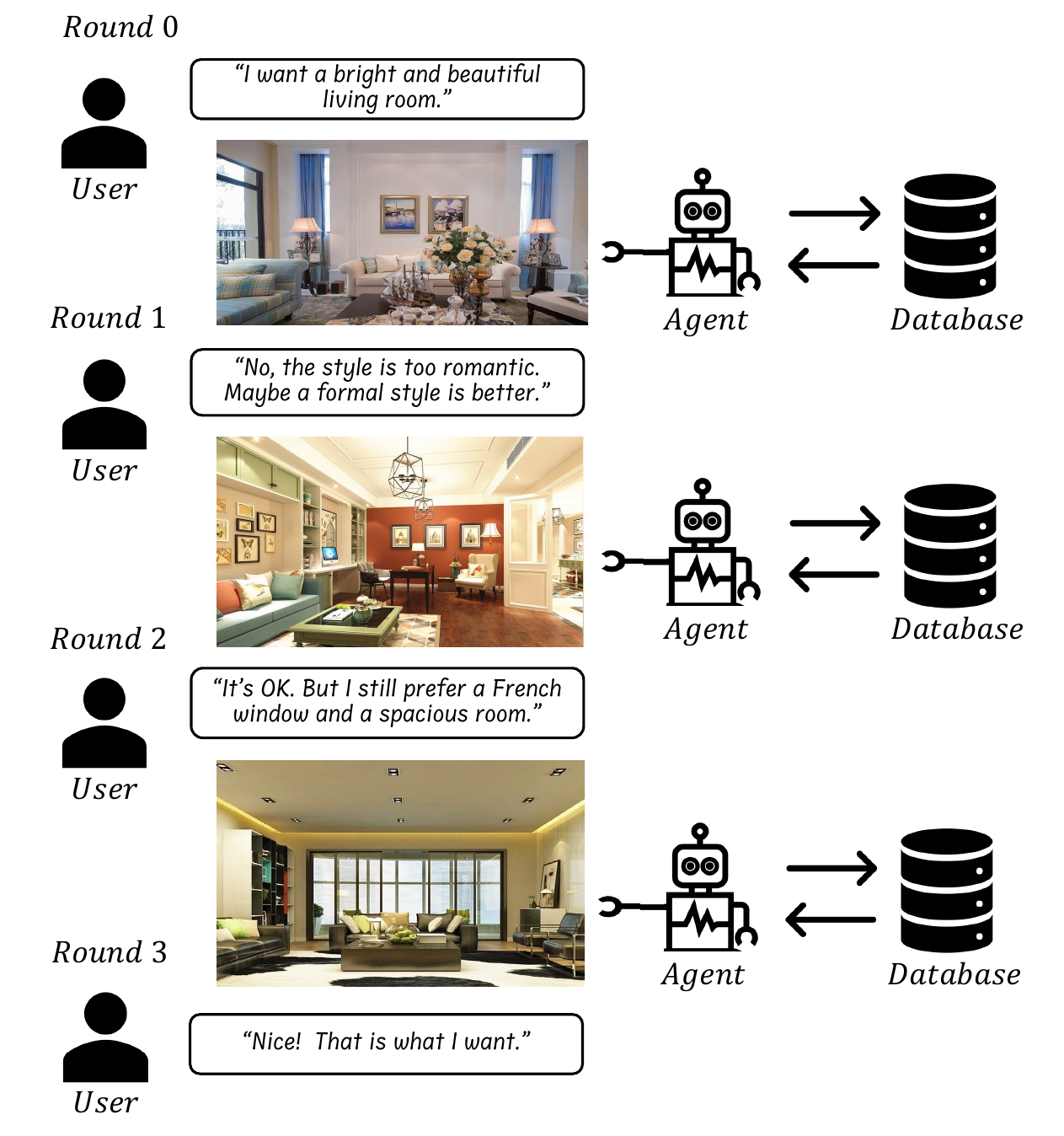}
  \vspace{-0.4cm}
  \caption{User-agent interaction process in IIDRS.}
  \label{fig:problemFormulation1}
  \vspace{-0.4cm}
\end{figure}

\section{RELATED WORK}

\subsection{Intelligent Interior Design Systems}
In the field of interior design, one direction is to learn furniture layouts for indoor scenes. For example, SceneFormer \cite{wang2021sceneformer} uses a self-attention mechanism to learn object relations. Di {\it et al.} \cite{di2021multi} use a multi-agent reinforcement learning-based scene design method to learn the optimal 3D furniture layout. With the development of deep generative models \cite{sharma2018chatpainter, gu2022vector, mildenhall2021nerf,ramesh2021zero}, indoor scene generation has become another important research direction. For example, Text2Room \cite{hollein2023text2room} generates 3D indoor scenes based on the input text. Text2Scene \cite{tan2019text2scene} recurrently generates objects and attributes in the current scene according to the input text. NeuRIS \cite{wang2022neuris} employs a neural rendering framework to handle texture-less areas and generate high-quality 3D indoor scenes from 2D images. Different from existing works, we focus on identifying a user's hidden preferences and choosing the proper initial design. We propose an interactive recommendation system to solve this problem.



\subsection{Interactive Recommendation}
Interactive recommender systems aim to extract user preferences through multiple rounds of interactions, and recommend items that match their interests, including attribute-based dialogue recommendation~ \cite{sun2018conversational,lei2020estimation,deng2021unified} and generative dialogue recommendation~ \cite{li2018towards,chen2019towards,zhou2020improving,zhou2020towards}. 
Traditional methods usually interact with users in a textual dialogue manner \cite{li2018towards}.
However, interactive recommendation in design scenarios depend on the user's intuitive visual experience.
Therefore, visually-grounded dialog system researches have also emerged in e-commerce platforms in recent years \cite{zhang2020reward,yuan2021conversational,delmas2022artemis,guo2018dialog}. For example, Yuan {\it et al.} \cite{yuan2021conversational} proposes a conversational fashion image retrieval method that predicts the desired image based on text and image information in the conversation history. 
Different existing works, we propose a method for decoration recommendation, which contains more complex attributes and involve more diversified multi-modal interaction than single items.

\section{PRELIMINARIES}

\subsection{Problem Formulation}
\label{sec:problemF}
The overall procedure for the interactive interior design recommendation process is shown in Figure \ref{fig:problemFormulation1}. The user interacts with the agent for multiple rounds. At the $t$-th round, the user provides textual requirements or feedback $f_t$ on the recommended decoration. The agent then recommends a decoration $a_t$ from the database $D$. The user determines the gap between the recommended an interior design case $a_t$ and the ideal case. Accordingly, they provide feedback $f_{t+1}$ and an overall satisfaction score $r_t$ on a five-level scale: \textit{strongly dislike}, \textit{dislike}, \textit{neutral}, \textit{like}, and \textit{strongly like}. We assume that there exists an ideal decoration $T$ that satisfies the user. The interaction ends either when the ideal decoration is recommended or when the maximum number of rounds $t_{max}$ is reached. Intuitively, even if the user is initially unclear about the ideal decoration, their feedback and satisfaction can be regarded as generated from the differences between the recommended decoration and their ideal decoration, which guide the exploration for the ideal decoration. Our objective is to train an agent with an effective strategy to recommend a design case $a_t$, using historical interactions $<f_0, a_0, r_0, f_1, a_1, r_1, \cdots, f_t>$, and to reach the ideal case $T$ within the fewest number of rounds.

\subsection{Data Description}
\label{sec:data}
\begin{figure}
  \includegraphics[width=0.4\textwidth]{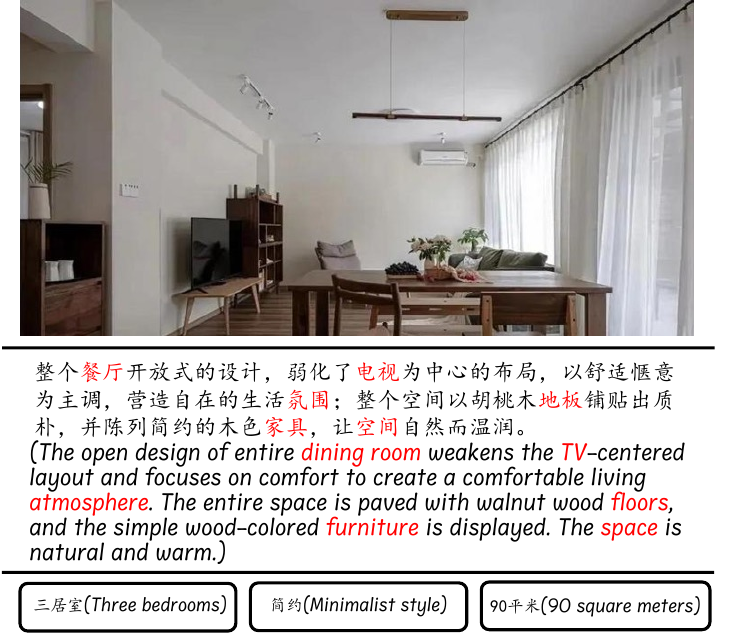}
  \vspace{-0.2cm}
  \caption{An interior design case from our dataset.}
  \label{fig:dataset1}
  \vspace{-0.4cm}
\end{figure}
The interior design data used in this paper were gathered from a popular Chinese interior design website\footnote{\url{https://www.jiajuol.com}}. The data comprises 90,000 decoration cases covering 20 different decoration styles and 10 room types. Each instance consists of images, descriptive texts, as well as tags such as house types and room styles. 
The text consists of lengthy paragraphs that include multiple sentences of subjective descriptions. These descriptions are provided by expert designers and focus on the decoration of a scene, including both overall descriptions of the entire scene, such as atmosphere and color, as well as specific views of individual objects, such as French windows and pendant lights. We create an object/attribute dictionary with 300 object types to identify different aspects of these descriptions. An illustrative example is shown in Figure \ref{fig:dataset1}, where the red words indicate objects/attributes. 
Formally, we formulate each interior design instance as $(d^j, l^j, \{o^j_i\}^{M_j}_{i=1})$, where $d^j$ denotes the decoration image, $l^j$ denotes the description, $\{o^j_i\}^{M_j}_{i=1}$ refers to objects set appearing in the $j$-$th$ instance's description, $M_j$ refers to the number of such objects. Notably, since textual descriptions may not always be available for online interior design platforms, we develop our system for the generic case where the database $D=\{d_j\}^N_{j=1}$ only contains images, where $N$ denotes the total number of images in the database.
The textual descriptions and objects are optional to develop a synthetic environment for training augmentation and model evaluation, which are detailed in Section~\ref{sec:fbGeneration} and Section~\ref{sec:usermodel}, respectively.

\section{Method}


\subsection{RL Formulation}
\label{sec:R}
We define the interactive interior design recommendation process as a Markov Decision Process (MDP) with the formulation $(\mathcal{S}, \mathcal{A}, \mathcal{R})$, where $\mathcal{S}$, $\mathcal{A}$, and $\mathcal{R}$ denote the state, action, and reward space, respectively. At round $t$, the state $s_t$ is formulated as 
\begin{equation}
    s_t = <f_0, a_0, f_1, a_1, \cdots, a_{t-1}, f_t>.
\end{equation}
In particular, $s_0=f_0$ is the first request from the user. We use $F_t=\{f_i\}_{i=0}^t$ to represent the historical feedback from the user and $A_t=\{a_i\}_{i=0}^{t-1}$ to represent the historical recommended images selected by the agent. The action $a_t \in \mathcal{A}$ (i.e., the recommended image) is selected from action space $D$ (i.e., the entire instance database) with a learned policy $\pi_\theta$ by a network parameterized $\theta$. To balance the exploration-exploitation trade-off during the training process, we involve a parameter $\epsilon$ valued from [0,1]. When $\epsilon=1$, the agent exploits the known optimal strategy without any exploration. When $\epsilon=0$, the agent explores the environment fully randomly, without considering any known information. 
We set the reward $r_t \in \mathcal{R}$ according to the five levels of user satisfaction score, labeled as $\{-2,-1,0,1,2\}$ respectively. Moreover, if the agent successfully recommends the user's ideal instance at round $t$, the reward $r_t$ is assigned a value of $10$ and the process is terminated.
The long-term return is formulated as discounted cumulative reward: 
\begin{equation}
    v_t = r_t + \gamma r_{t+1} + \gamma^2 r_{t+2} + \dots = {\textstyle\sum_{k=0}^{t_{max}-t} \gamma^k r_{t+k}},
\label{eq:vt}
\end{equation}
where $\gamma$ represents the discount factor.
The objective of the agent is to learn an optimal policy $\pi_\theta$ through interacting with the environment by maximizing $v_t$. 
Notably, different from existing works  \cite{guo2018dialog}, our supervision signal does not come from the rank of a specific target item since the ideal instance is defined to be implicit in our problem formulation, which is unknown unless the agent succeeds in finding it in $t_{max}$ rounds.

\begin{figure*}
  \includegraphics[width=0.95\textwidth]{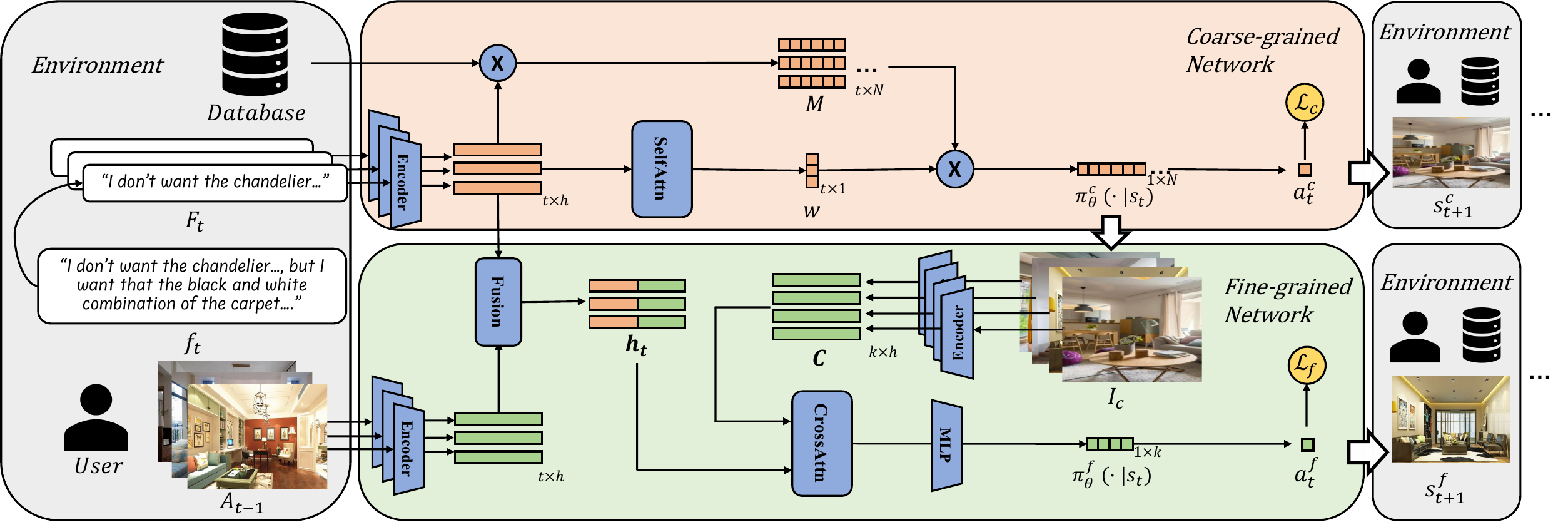}
  \vspace{-0.2cm}
  \caption{Architecture of DecorRCFN.}
  \label{fig:Recommendation.png}
  \vspace{-0.4cm}
\end{figure*}
\subsection{Decoration Recommendation Coarse-to-Fine Policy Network}
In the decoration recommendation problem, the action space is large due to the many possible decorations to recommend. Learning a fine-grained policy on all possible choices with a complex model can lead to low exploration-efficiency and make model convergence difficult. Inspired by the classic preranking-ranking 2-step recommendation schema, we propose DecorRCFN, a coarse-to-fine policy network consisting of two sub-policy modules. Its architecture is shown in Figure~\ref{fig:Recommendation.png}. Specifically, the coarse-grained policy network learns the policy $\pi_\theta^c$ for selecting a set of candidate actions from the entire image set $D$, based on the similarity of user feedback and decoration representations. The fine-grained policy network then models multimodal state-action interaction and learns the policy $\pi_\theta^f$ for selecting the final action from the candidate images.

\subsubsection{Coarse-grained Policy Network}
\label{sec:CGN}
Given the large action space, it is necessary to reduce the number of candidate interior design cases from the entire database $D$ that need to be considered by a complex, fine-grained policy. Intuitively, the historical feedback provided by users can be seen as incomplete descriptions of desired decoration. Therefore, a straightforward idea is to extract a representation of the user's demands from their historical feedback and match it with representations of decoration images. 
We employ CLIP \cite{radford2021learning, yang2022chinese}, a prominent multimodal pre-trained model for image-text matching, as the encoder module for generating embeddings $\boldsymbol{F_t} \in \mathbb{R}^{t \times h}$ and $\boldsymbol{D_t} \in \mathbb{R}^{N \times h}$ for textual feedback $F_t$ and decoration images in $D$:
\begin{equation}
\begin{aligned}
\boldsymbol{F_t} &= \{\text{Encoder}(f_i) | f_i \in F_t\},\\
\boldsymbol{D} &= \{\text{Encoder}(d_j) | f_j \in D\},
\end{aligned}
\end{equation}
where $h$ represents the embedding dimension. 
Then, we calculate a similarity matrix $M \in \mathbb{R}^{t \times N}$, the element $M_{i,j}$ represents the similarity between historical feedback $f_i \in F_t$ and $d_j \in D$ can be computed as:
\begin{equation}
M_{i,j} = \text{sim}(\text{Encoder}(f_i), \text{Encoder}(d_j))
\end{equation}
where $\text{sim}(\cdot, \cdot)$ computes the cosine similarity of two embeddings, outputting a value ranging from $-1$ to $1$. Based on the similarity matrix, a trivial candidate selection method is to add up the scores of all the feedback and retrieve the images with the highest scores. However, this approach may be suboptimal because different feedback contains various information and may have differing levels of importance in reflecting user preferences.
Therefore, we consider candidate selection as a learnable policy, with the action space defined as the entire interior design database, i.e., $\mathcal{A}_c=D$.
To model the importance of each $f_i \in F_t$, we use a multi-head self-attention module \cite{vaswani2017attention}, formulated as 
\begin{equation}
\begin{aligned}
    \text{SelfAttn}(X) = \text{softmax}(\frac{(XW_Q)(XW_K)^T}{\sqrt{d_k}})(XW_V)
\end{aligned}
\end{equation}
where $X$ represents the input for the self-attention layer, $W_Q$, $W_K$, and $W_V$ are trainable parameters, $d_k$ is the key dimension in the attention mechanism. Based on the self-attention module, we get the attention vector $w \in \mathbb{R}^{t}$ as $w=\text{MLP}(\text{SelfAttn}(\boldsymbol{F_t}))$, where MLP is a Multi-Layer
Perceptron, and accordingly calculate the selection probability as 
\begin{equation}
\begin{aligned}
\pi^c_\theta (\cdot|s_t) = \text{softmax}(w^TM). 
\end{aligned}
\end{equation}
Since direct supervision of the proper candidate is unavailable, we train the coarse-grained policy with a reinforcement learning schema. Specifically, we employ coarse-grained policy to draw the recommendation sequence and adopt Policy Gradient (PG)  \cite{williams1992simple} to train the model. For each step, among all the images in $D$, the one with the largest sore is selected as the action, formulated as $a_t^c = \arg\max_{a^c \in \mathcal{A}_c}\pi^c_\theta(a^c|s_t^c)$. 
The agent aims to maximize the object function $J_c$, formulated as: 
\begin{equation}
\begin{aligned}
    J_c &= \sum_{a^c \in \mathcal{A}_c } \pi_\theta^c(a^c|s^c) \cdot v, \\
    \nabla_\theta J_c &\approx \sum_{t}\nabla_\theta log(\pi_\theta^c(a^c_t|s_t^c)) \cdot v_t, \\
\end{aligned}
\end{equation}
where $\pi_c(a^c|s)$ denotes the coarse-grained policy. The optimization problem can be reformulated as the minimization of the loss function $\mathcal{L}_c$, which is given by:
\begin{equation}
    \mathcal{L}_c = -\sum^{t_{max}}_{t=1} log(\pi_\theta ^c(a_t^c|s_t^c)) \cdot v_t.
    \label{eq:l1}
\end{equation}
Notably, the similarity calculation requires encoding images for all the images in the database everytime the parameters are updated. To reduce model complexity, we freeze the image embeddings in CLIP module so that we can pre-calculate them for quick retrieval. 

\subsubsection{Fine-grained Policy Network}
The coarse-grained network usually prioritizes the textual modality but ignores how it interacts with the image modality in recommendation history. Indeed, since each textual feedback specifies user opinion for specific visual characteristics of the given image, looking into the informative visual signals based on the feedback can facilitate user preference modeling. Therefore, instead of directly choosing the final action with coarse-grained policy, we select the top-$k$ decorations as candidate set $I_c$.
Then, we train the fine-grained policy network in DecorRCFN to further determine the best choice from the candidate set. 
Specifically, the fine-grained policy network models the semantic interaction between $A_{t-1}$, $F_t$, and each candidate in $I_c$.
We obtain embeddings $\boldsymbol{A_{t-1}}$ of the action history $A_{t-1}$ and $\boldsymbol{C}$ of the candidate images $I_c$ through the CLIP encoder module:
\begin{equation}
\begin{aligned}
    \boldsymbol{A_{t-1}} &= \{\text{Encoder}(a_i)|{a_i \in A_{t-1}}\}, \\
    \boldsymbol{C} &= \{\text{Encoder}(c_i)|{c_i \in I_c}\}.  \\
\end{aligned}
\end{equation}
To capture user preferences from multiple modalities in multiple rounds, we fuse the embeddings of feedback sentences $\boldsymbol{F_t}$ and the action history $\boldsymbol{A_{t-1}}$ to form a new embedding vector $\boldsymbol{h_t}$. 
In order to select the action of the fine-grained network that best matches the user preferences, we use a multi-head cross-attention module \cite{guo2022context,wei2020multi} to compare the $\boldsymbol{h_t}$ and $\boldsymbol{C}$. The cross-attention module is defined as:
\begin{equation}
    \text{CrossAttn}(X_Q,X) = \text{softmax}(\frac{(X_QW_Q) (XW_K)^T }{\sqrt{d_k}}) (XW_V), 
    \label{eq:crossattn}
 \end{equation}
where $X_Q$ and $X$ are the query and key-value matrices, respectively.
We utilize $\boldsymbol{h_t}$ as the query matrix and $\boldsymbol{C}$ as the key-value matrix to obtain the attention matrix, which is then fed into a Multi-Layer Perceptron (MLP) with Softmax activation to obtain the probability of each actions:
\begin{equation}
\begin{aligned}
    \pi^f_\theta(\cdot|s_t^f) = \text{softmax}(\text{MLP}(\text{CrossAttn}(\boldsymbol{C},\boldsymbol{h_t}))). \\
\end{aligned}
\end{equation}
Therefore, the optimal action of the fine-grained network is represented as $a_t^f = \arg\max_{a^f \in \mathcal{A}_f} \pi^f_\theta(a^f|s_t^f)$, where $\mathcal{A}_f = I_c$ denotes the action space (i.e., candidate pool) of the network.
Similar to the coarse-grained policy, Based on PG algorithm, we define the object function and loss function of the fine-grained network as follows:
\begin{equation}
\begin{aligned}
\mathcal{L}_f &= -\sum_{t} log(\pi_\theta^f (a_t^f|s_t^f)) \cdot v_t. \\
\end{aligned}
\end{equation}
We combine the coarse-grained and fine-grained loss to formulate the overall Coarse-to-fine Network loss, based on sampled recommendation paths from the final policy $\pi_\theta=(\pi_\theta^c, \pi^f_\theta)$, as
\begin{equation}
\begin{aligned}
\mathcal{L}_{cf} =- \sum^{t_{max}}_{t=1}(\alpha \cdot log(\pi^c_\theta (a_t^f|s_t^f)) + log(\pi_\theta^f (a_t^f|s_t^f))  ) \cdot v_t,
\end{aligned}
\label{eq:Loss}
\end{equation}
where $\alpha \in [0, 1]$ balances the importance of the two policies.

\subsection{Training Augmentation with Object-aware Feedback Generation}
\label{sec:fbGeneration}
\begin{figure}
  \includegraphics[width=0.4\textwidth]{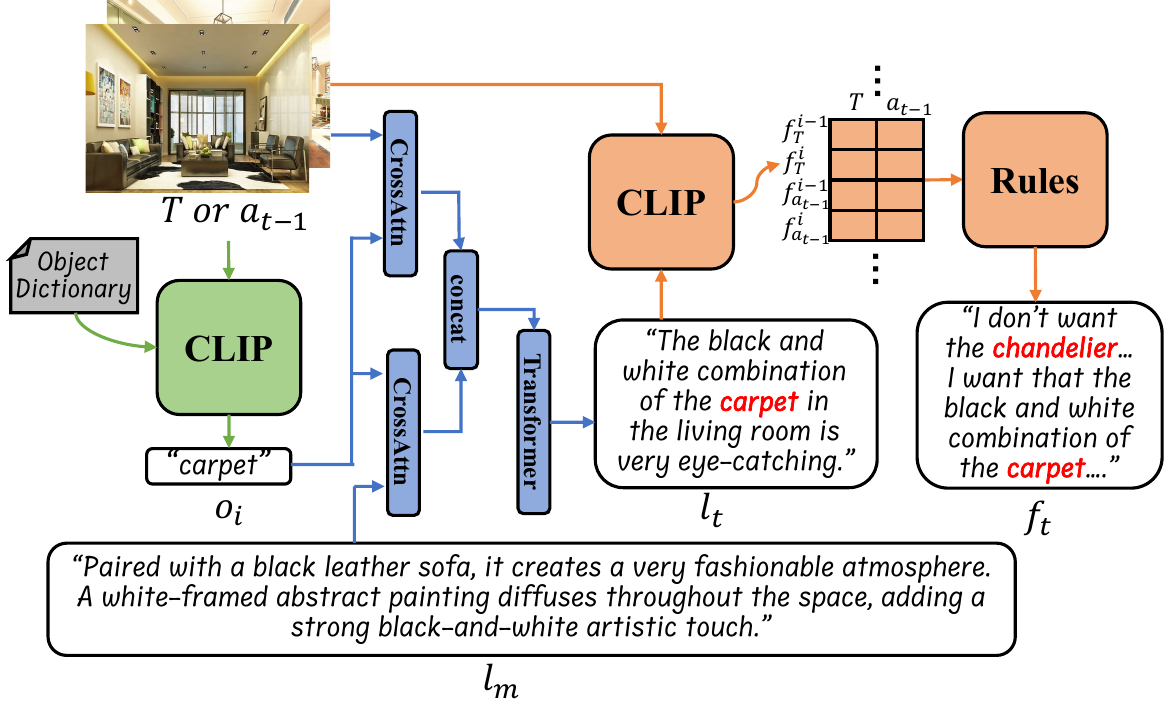}
  \vspace{-0.3cm}
  \caption{Framework of Object-aware Feedback Generation.}
  \label{fig:UserModel1}
  \vspace{-0.4cm}
\end{figure}

Real-world user feedback exhibits subjectivity, imprecision, uncertainty, and diversity nature. The model needs abundant explorations before it learns effective strategies to dig the preference from the complex and uncertain user feedbacks whose patterns vary with user's personal habits. However, collecting user feedback through real interaction can be expensive. The limited data can harm model performance for easily causing overfiting problem. Therefore, data augmentation is important to enhance model generalization in complex online scenarios. To enhance model generalization in complex online scenarios, we propose augmenting our training procedure with diverse simulated feedback dynamically generated based on the differences between the recommended decoration and the target decoration. As a decoration design can be described from various aspects \cite{hossain2019comprehensive, deng2021transvg, wang2022ofa}, our method generates difference descriptions in an objects/attributes-aware manner, containing three steps: object detection, object-aware image captioning, and difference description filtering. The overall framework is shown in Figure \ref{fig:UserModel1}.


\subsubsection{Object Detection}
\label{sec:objectDetection}

Our dataset includes selective descriptions of objects and attributes that appear in an image. We first train a model to identify objects or attributes that can be found in the image. Specifically, we treat the object as text and fine-tune a CLIP model with image-object pairs. This allows the model to calculate the similarity between an image $d$ and an object $o$. A higher similarity score indicates a higher chance of the object existing in the image.

\subsubsection{Object-aware Image Captioning}

After recognizing objects in an image, we train an object-aware image captioning model to generate captions describing these objects in their corresponding images.
To be specific, based on our decoration dataset, we build a training set in which each sample formulated as $<(d,o,l_m), y>$, where $d$ represent the image, $o$ represents an object, $y$ denotes the target sentence that describes object $o$ within image $d$, $l_m$ is the rest descriptions after masking the target sentence. 
To model these data, our model embeds the image $d$ with a CNN backbone, and embed the text $o_i$ and $l_m$ with a BERT \cite{devlin2018bert} encoder. The embeddings are represented as $\boldsymbol{d}$, $\boldsymbol{o}$, and $\boldsymbol{l_m}$.
Then, we utilize a transformer module, which has proven to be superior in image captioning tasks \cite{vaswani2017attention,stefanini2022show}, to generate textual descriptions. 
However, directly feeding all types of embeddings into the transformer can lead to high computation costs due to excessive length of sequences. Therefore, we involve two cross-attention modules (as Eq. \ref{eq:crossattn}) to reduce the sequence length. The two cross-attention modules both use $\boldsymbol{o}$ as $Q$ and $\boldsymbol{d}$ and $\boldsymbol{l_m}$ as $K$ and $V$, respectively. The features contain information of the object within the images and descriptions.
Finally, we concatenate the features and feed them into the transformer module to produce an object-aware image caption $\hat{y}$. Following existing image captioning works, we train the model by minimizing cross-entropy loss between target description $y$ and the generated sentence $\hat{y}$. 

\subsubsection{Difference Description Filtering}
So far, a variety of sentences can be generated to describe different aspects of a decoration image. To filter the sentences that describe differences between the current image and target image, we developed a filter module with a CLIP model fine-tuned on image-sentence pairs in our dataset. Using the filter, we identify sentences that have high similarity to one image and low similarity to the other as describing their differences. Ultimately, we synthesize user behavior based on the rules introduced in Section \ref{sec:usermodel}. This enriches the diversity of feedback and improves the recommendation performance.

\begin{table*}[]
\centering
\begin{tabular}{@{}lllllllllllll@{}}
\toprule
\multirow{3}{*}{\textbf{Model}} &
  \multicolumn{6}{c}{\textbf{\begin{tabular}[c]{@{}c@{}}Feedback from \\ Generated Sentences(for test)\end{tabular}}} &
  \multicolumn{6}{c}{\textbf{\begin{tabular}[c]{@{}c@{}}Feedback from\\ Real-world Sentences(for test)\end{tabular}}} \\ \cmidrule(l){2-7}
  \cmidrule(l){8-13} 
 &
  \multicolumn{2}{c}{Max round = 10} &
  \multicolumn{2}{c}{Max round = 8} &
  \multicolumn{2}{c}{Max round = 6} &
  \multicolumn{2}{c}{Max round = 10} &
  \multicolumn{2}{c}{Max round = 8} &
  \multicolumn{2}{c}{Max round = 6} \\
 & r@1      & r@2 & r@1         & r@2 & r@1         & r@2 & r@1         & r@2 & r@1         & r@2 & r@1         & r@2 \\ \midrule
Random       & 0.43\% & 0.53\% & 0.35\% & 0.40\% & 0.31\%   & 0.42\%      & 0.60\%  & 0.71\%     & 0.35\%   &  0.42\%    & 0.31\%   & 0.39\% \\
CLIP-greedy          & 42.23\%          &  47.05\%    & 33.15\%        & 38.86\%     & 21.84\%          & 27.30\%     & 28.12\%          & 32.78\%     & 25.46\%          &  28.64\%    & 21.70\%          & 24.43\%     \\
CLIP-random     
&44.11\%&48.97\%&32.80\%&38.97\%&21.27\%&28.82\%&32.41\%&33.86\%&27.80\%&30.03\%&22.62\%&25.67\% \\
DecorRCGN       & 48.12\%          &  54.46\%   & 38.04\%          &  44.14\%    & 21.66\%          & 27.16\%     & 32.26\%          & 33.79\%     & 28.33\%          & 30.98\%      & 21.66\%          & 24.44\%     \\
DecorRCFN  & \textbf{49.82\%} &  \textbf{55.56}\%    & \textbf{38.23\%} &  \textbf{44.64}\%    & \textbf{24.53\%} & \textbf{30.88}\%     & \textbf{33.80\%} & \textbf{35.63\%}     & \textbf{29.40\%} & \textbf{31.76\%}     & \textbf{23.57\%} &  \textbf{26.27\%}    \\ \bottomrule
\end{tabular}
\vspace{0.2cm}
\caption{Performance Evaluation for IIDRS.}
\label{tab:cidrs}
\vspace{-0.6cm}
\end{table*}

\section{Experiment}

\subsection{Experimental Setup}
\subsubsection{Datasets and Settings}
To ensure both the effectiveness and efficiency of reinforcement learning training in a large action space,, we further selectively screen detailed and clear interior design cases from the data. Specifically, we select 5731 samples from the screened data to be used as the experimental dataset for subsequent analysis.
In order to more accurately evaluate the performance and generalization of the model, we split the dataset into training and testing sets in a 1:1 ratio, which ensures that the size of the action space is the same. The CLIP models have been fine-tune with our training set.
To demonstrate the effectiveness of Object-aware Feedback Generation, we then create two datasets based on the source of feedback: Real-world Sentences(RS), which consists of synthesized feedback derived from real-world descriptions, and Generation Sentences(GS), which utilizes feedback generated through the Object-aware Feedback Generation model.
We select 10 objects with the highest similarity to each image. 
The parameters of DecorRCFN are empirically set as: $\epsilon=0.8$ (Section \ref{sec:R}), $\gamma=0.8$ (Equation \ref{eq:vt}), $\alpha=0.1$ (Equation \ref{eq:Loss}), the number of candidates $k=4$, and the learning rate $lr=10^{-5}$.

\subsubsection{User Simulation}
\label{sec:usermodel}
For multi-round interactive recommendation systems, the environments needs real users to converse with the agent and provide rewards, which is challenging to create for research purposes \cite{DBLP:journals/corr/DhingraLLGCAD16}. A common solution is using simulated users to interact with the agent. In IIDRS, the user simulator simulates the ideal case by randomly drawing a decoration image from the database. Accordingly, it simulates the following three behaviors: 1) providing feedback $f_t$ according to the given image $a_{t-1}$ and the ideal image $T$; 2) providing satisfaction level (Section \ref{sec:R}); 3) terminating the interaction when the target case $T$ is recommended or the number of rounds exceeds $t_{max}$. To simulate feedback, we create rules that use a set of description sentences for the current and target images to construct candidate feedback. We evaluate the feedback using the CLIP model to determine whether a given sentence accurately describes the differences between the two images. An effective feedback should exhibit a significant contrast between the similarity score to $T$ and the $a_t$. If a sentence $f$ has a high similarity score to $a_t$ but a low similarity score to $T$, we use the template "I don't like \{$f$\}" to generate a sentence describing the user's dislike of the current image. Conversely, if $f$ has a high similarity score to $T$ but a low similarity score to $a_t$, we use the template "I prefer \{$f$\}" to generate a sentence describing the user's demand for the ideal decoration. During the interaction, feedback sentences are randomly sampled from the constructed candidates based on the current and target image. For satisfaction level simulation, we design a rule based on the rank $\text{rank}(a_t, T, D)$ of $a_t$ in terms of its embeddings' cosine similarity to $T$ among images in $D$. Specifically, the satisfaction level $r_t$ is defined as
\begin{equation}
\begin{aligned}
    r_t = 
    \left \{  
    \begin{array}{llll}
    2, &   & \text{rank}(a_t,T,D) \leq l_0\\
    1, &   & l_0 < \text{rank}(a_t,T,D) \leq l_1\\
    0, &   & l_1 < \text{rank}(a_t,T,D) \leq l_2\\
    -1, &   & l_2 < \text{rank}(a_t,T,D) \leq l_3\\
    -2, &   & \text{rank}(a_t,T,D) > l_3,
    \end{array}
    \right.
\end{aligned}
\end{equation}
where $l_{\{0,1,2,3\}}$ denotes the thresholds for different satisfaction levels. We set $l_{\{0,1,2,3\}} = \{10,20,30,50\}$ in our experiment.

\subsection{Recommendation Performance Evaluation}

\subsubsection{Baseline Performance Analysis.} We primarily evaluate the effectiveness of our recommendation model using recall@1($r@1$) and recall@2($r@2$). As online reinforcement learning for multi-round text-image recommendation represents a new task, we compared our approach with two baseline methods: Random search(Random) and CLIP models(CLIP). 
Specifically, for Random search, we randomly select images from the database for $Max round$ rounds. For CLIP-greedy model, we select the image that receive the highest similarity score with all feedback during an interaction, calculated by a fine-tuned CLIP model. 
However, the greedy strategy is not always optimal, as user preferences are uncertain during the whole recommendation process.
Therefore, we introduce a CLIP-random baseline, which is based on CLIP but randomly selects the next image from the top 4 candidates calculated by CLIP.

Table \ref{tab:cidrs} presents our experimental results.
Random search performed poorly, failing to hit the target image. On the other hand, the fine-tuned CLIP model achieved relatively good performance.
We observed that the score of $r@2$ using CLIP-greedy outperformed $r@1$ by almost 5\%, indicating that the image with the highest similarity score is not always the optimal choice. It also reflects that the user feedback is information limited, leading to the most similar image not necessarily being the most appropriate choice.
Furthermore, the random strategy appears to be less stable and uncontrollable in performance compared to Greedy search and other models. Specifically, the variance of the results obtained using CLIP-random is higher than that of CLIP-greedy and other methods.

\subsubsection{DecorRCGN and DecorRCFN}

In Table \ref{tab:cidrs}, the Decoration Recommendation Coarse-Grained policy Network (DecorRCGN) and the Decoration Recommendation Coarse-to-Fine policy Network (DecorRCFN) exhibit superior performance compared to the baselines in general.
DecorRCGN ranks the actions in database by recalculating the weights of different feedback sentences(Section \ref{sec:CGN}). To illustrate its effect, we conduct an experiment as shown in Table \ref{tab:weights}. While all feedback sentences carry equal weights in CLIP-based models, DecorRCGN assigns different weights to each sentence based on its content. For instance, the first feedback `White' being the shortest and providing the least information, receives the lowest weight of 0.5557.
In contrast, the third line sentence, `A white bedroom with enough space for kids to play', containing the most information, obtains the highest weight of 0.9711.
After assigning weights to each feedback sentence based on their importance, the scores of similarity between each image and sentence are summed up. When selecting the next action in DecorRCGN, it is found that the third sentence, which contains the most information, has the highest weight and thus contributes the most to the next image.
\begin{table}[]
\begin{tabular}{@{}lll@{}}
\toprule
\textbf{Feedback in English}                     & \textbf{CLIP} & \textbf{DecorRCGN} \\ \midrule
White                                            & 1             & 0.5557              \\
A white bedroom                                    & 1             & 0.8720              \\
\begin{tabular}[c]{@{}l@{}}Spacious white bedroom for kids to play.\end{tabular} & 1             & 0.9711              \\ \bottomrule
\end{tabular}
\caption{Weights of different feedback sentences.}
\label{tab:weights}
\vspace{-1cm}
\end{table}
As shown in Table \ref{tab:cidrs},  DecorRCGN outperforms both CLIP-greedy and CLIP-random by 5.8\% and 4.0\%, respectively, in terms of $r@1$ when $Max round=10$. 
Nonetheless, DecorRCGN has two drawbacks. Firstly, it still employ the greedy strategy when selecting the next action, ignoring the fact that the image with the highest score does not necessarily represent the best choice. Secondly, when the $Max round$ is set to $6$, DecorRCGN fails to surpass CLIP-greedy due to insufficient feedback information. However, our proposed DecorRCFN effectively addresses these issues.
The DecorRCFN model employs a more refined policy by selecting the top k candidates filtered by DecorRCGN. This helps to avoid always selecting images with the highest scores, which may not necessarily be the optimal choice.
Furthermore, DecorRCFN enhances candidate comparison by utilizing information from both feedback history and images recommended in the previous round. This results in improved performance of DecorRCFN, especially achieving 24.53\% in terms of $r@1$ when textual modality is limited when $Max round=6$, shown in Table \ref{tab:cidrs}. In both the test set with generated sentences(GS) and the test set with real-world sentences(RS), DecorRCFN consistently outperforms DecorRCGN and achieves the best performance.

\subsubsection{Ablation Study.}
We conduct an ablation study to illustrate the effectiveness of the RL compared to supervised learning. In this problem formulation, the agent is unable to acquire additional information beyond feedback and rewards. Therefore, the loss of supervised learning can only incorporate rewards.
In Table \ref{tab:reward}, $\mathcal{L}_{c}$ and $\mathcal{L}_{cf}$ represent the losses of the DecorRCGN and DecorRCFN, respectively. 
The $reward$ indicates that the loss is calculated solely based on $r_t$, which can be regarded as reward supervised learning. 
The $value$ means the loss is calculated according to $v_t$ Equation \ref{eq:vt}, where $\gamma=0.8$ to get the best results. We compare the recall@1 performance of these settings with $Max round=10$. And the results indicate that models based on reward supervised learning achieve much lower values of $r@1$ compared to RL models. 
This is because when $\gamma=0$, the models never consider future rewards but only focus on the current choices, leading to performance similar to CLIP-greedy. Once we incorporate future rewards into the loss, the agent could choose a better action based on the entire episode. 

\begin{figure}
  \includegraphics[width=0.4\textwidth]{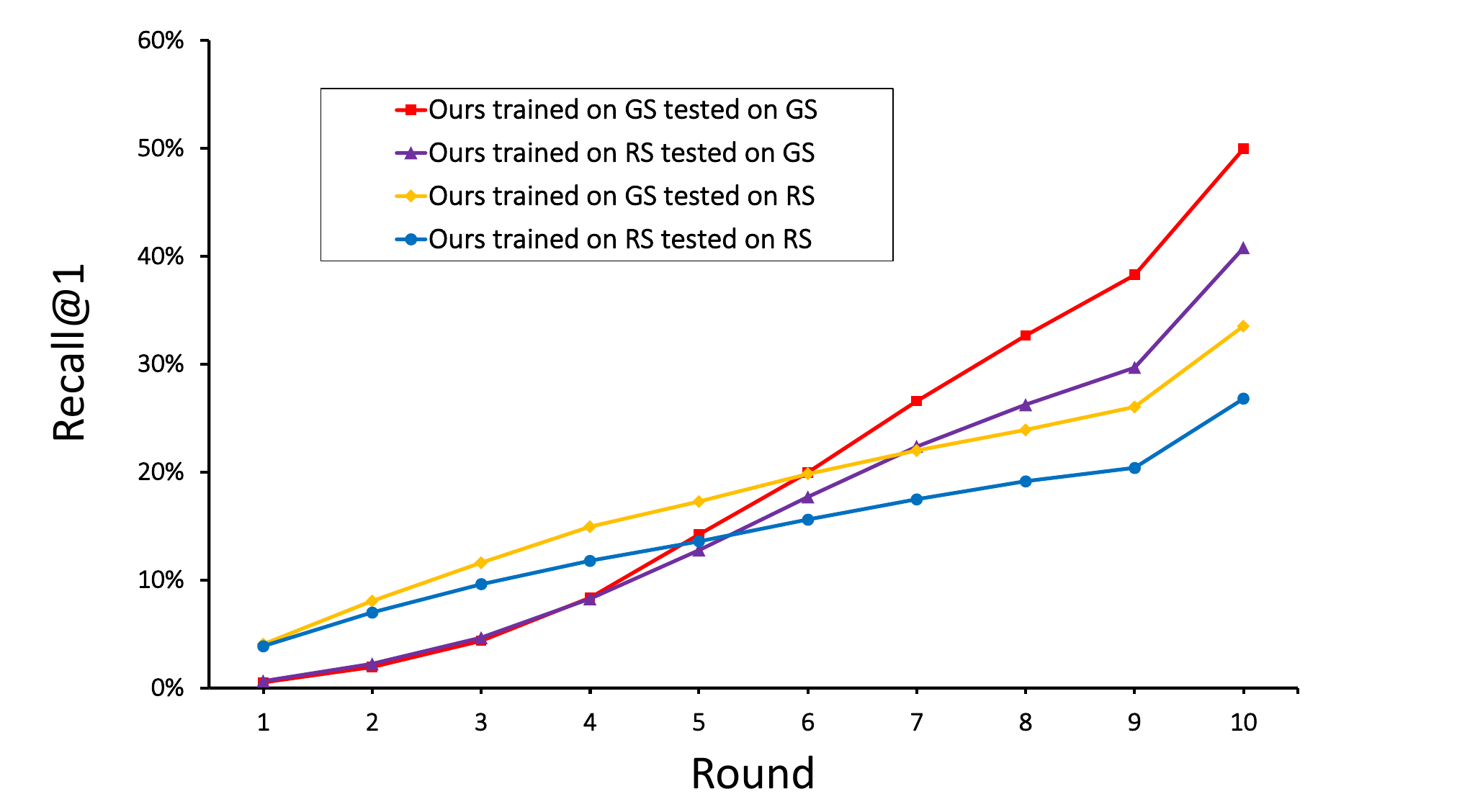}
  \vspace{-0.45cm}
  \caption{Influence of Object-aware Feedback Generation.}
  \label{fig:tableEachround}
  \vspace{-0.5cm}
\end{figure}

We conduct another ablation study on the influence of object-aware feedback generation.
Figure \ref{fig:tableEachround} displays the correlation between the number of rounds and the value of recall@1 when DecorRCFN is trained on different dateset and test on different dataset. Specifically, GS denotes the dataset using generated sentences from object-aware feedback generation, and RS denotes the dataset with real-world sentences. 
All of the curves show an upward trend. Specifically, the red and purple curves, which represent DecorRCFN tested on GS, exhibit increasingly steeper slopes as the number of rounds increases. This observation suggests that DecorRCFN's performance improves with additional feedback and image information from multiple rounds.
However, the models tested on RS(the yellow and blue curves) show slower growth. This is because the information from real-world sentences doesn't increase necessarily as the rounds increasing, due to the limited number of sentences in real-world descriptions.
Furthermore, the figure indicates that the models trained on GS(represented by the red and yellow curves) perform better than those trained on RS(represented by the purple and blue curves) in both test sets. This highlights the effectiveness of our object-aware feedback generation model (see Section \ref{sec:fbGeneration}), which enriches the diversity of user feedback, augments the data in the training set, and ultimately improves the model performance.

\begin{table}[]
\begin{tabular}{ccccc}
\toprule
\textbf{\begin{tabular}[c]{@{}c@{}}$\mathcal{L}_{c}$\\ (reward)\end{tabular}} &
  \textbf{\begin{tabular}[c]{@{}c@{}}$\mathcal{L}_{c}$\\ (value)\end{tabular}} &
  \textbf{\begin{tabular}[c]{@{}c@{}}$\mathcal{L}_{cf}$\\ (reward)\end{tabular}} &
  \textbf{\begin{tabular}[c]{@{}c@{}}$\mathcal{L}_{cf}$\\ (value)\end{tabular}} &
  \textbf{\begin{tabular}[c]{@{}c@{}}Max round=10\\recall@1\end{tabular}} \\ \midrule
$\checkmark$ &   &   &   & 42.43\%          \\
  & $\checkmark$ &   &   & 48.26\% \\
 &   & $\checkmark$ &   & 42.72\%          \\
  &  &   & $\checkmark$ & \textbf{49.82\%} \\ \bottomrule
\end{tabular}
\caption{Comparison between ours and reward supervised learning.}
\label{tab:reward}
\vspace{-0.5cm}
\end{table}

\begin{figure*}
  \includegraphics[width=0.78 \textwidth]{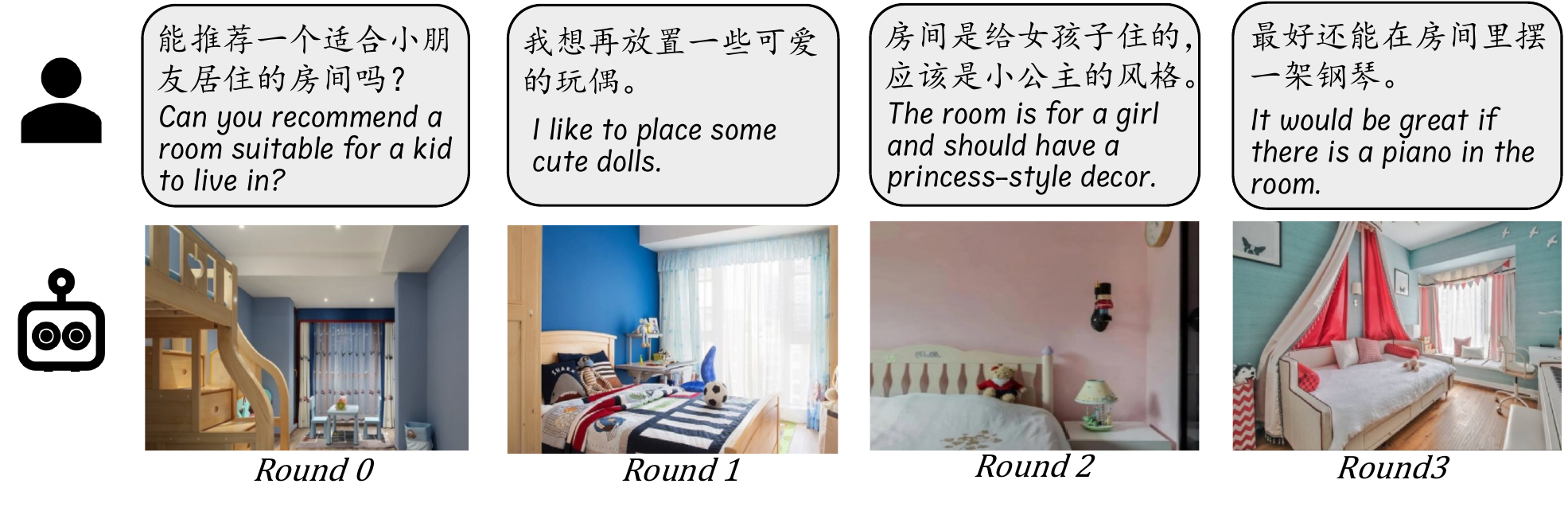}
  \vspace{-0.5cm}
  \caption{Real-world examples of interactions with our IIDRS.}
  \label{fig:example1}
\end{figure*}
\subsection{Object Description Model Evaluation}

We assess the object-aware feedback generation model using widely adopted metrics such as BLEU-1 and ROUGE-L \cite{hossain2019comprehensive}, to evaluate the fluency of the generated sentences. Additionally, to demonstrate that the model could produce high-quality sentences, we compare the similarity scores, CLIP-S, of the images and generated sentences.
Our generation model consists of the two cross-attention modules, called Double CrossAttn in Table \ref{tabel:des_model}. We compare it with the surely transformer module without cross-attention modules previously, shown in the first row of the table. 
Without cross-attention modules before, it joints the embedings of $d$, $o$ and $l_m$ together, leading to an increased sequence length. 
Consequently, Double CrossAttn achieves comparable results to Transformer, while also requiring less time for sentence generation due to shorter sequence length. Specifically, Double CrossAttn has an average generation time of $95.5ms$ per sentence, compared to Transformer's average of $218.8ms$.
To demonstrate the significance of the information from the masked descriptions $l_m$, we train a Single CrossAttn module without integrating the modality of $l_m$. 
The experimental results validate that the information from $l_m$ cannot be disregarded. The BLEU-1 and ROUGE-L scores of Single CrossAttn are much lower than those of Double CrossAttn, indicating that the sentences generated by Single CrossAttn are less fluent as than those generated by Double CrossAttn. Meanwhile, the CLIP-S of Double CrossAttn surpasses the score of Single CrossAttn by its 22\%. This implies that the sentences generated by Single CrossAttn are unable to pass the filter.
We additionally involve a temperature parameter(T), which increases the probability to generate different words. Finally, the performance of Double CrossAttn(T) surpasses Double CrossAttn in terms of BLEU-1 and ROUGE-L scores, though it achieves slightly lower scores for the CLIP-S.

\begin{table}[]
\begin{tabular}{@{}lllll@{}}
\toprule

\textbf{Module}           & \textbf{BLEU-1}  & \textbf{ROUGE-L} & \textbf{CLIP-S} \\ \midrule
Transformer              & 0.4396                    & 0.2423           & 27.76   \\
Single CrossAttn           & 0.2721                   & 0.0905           & 22.75  \\
Double CrossAttn                & 0.4317           & 0.2506           & \textbf{27.85}\\
Double CrossAttn(T)           & \textbf{0.4405}          & \textbf{0.2544}  & 27.67 \\
\bottomrule
\end{tabular}
\vspace{0.1cm}
\caption{Performance Evaluation for Object Description.}
\label{tabel:des_model}
\vspace{-0.8cm}
\end{table}

\subsection{User Study}
To demonstrate the effectiveness of our system in interacting with humans, we recruited 20 individuals with diverse backgrounds to test our system. Each user communicated with three models: DecorRCGN, DecorRCFN, and CLIP-greedy, without being informed which model they were interacting with during the entire process. We recorded each user's level of satisfaction after each round of interaction. Once a user was satisfied with the recommended image and decided to stop further recommendations, the model received a reward of 3 in both the current and subsequent rounds.
\subsubsection{Case Study}
An example interaction with our IIDRS is shown in Figure \ref{fig:example1}. The user cannot exactly describe their whole preferences at the beginning. For instance, they just ask for a room but don't require the piano at first round. 
As the number of rounds increases, the interior decoration cases provided by the agent guide the user to offer more specific feedback, which is also the reason why multi-round interactive recommendation systems outperform single-round ones. In Figure \ref{fig:example1}, our agent effectively satisfies both current and historical user needs.

\subsubsection{Experiments Evaluation}
\begin{figure}
  \includegraphics[width=0.4\textwidth]{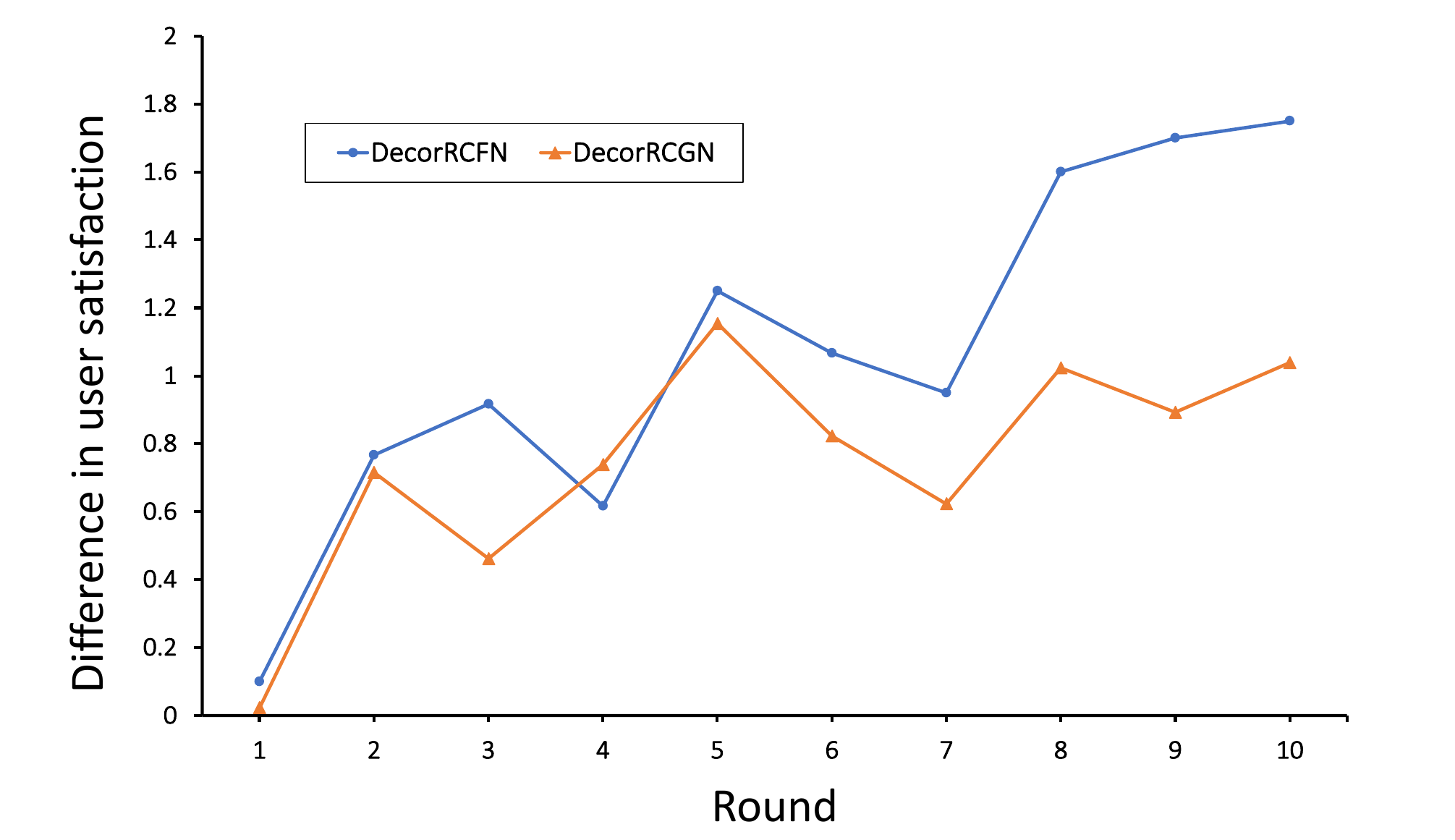}
  \vspace{-0.45cm}
  \caption{Differences in user satisfaction levels with different models compared to CLIP-greedy.}
  \label{fig:userStudy}
  \vspace{-0.5cm}
\end{figure}

We compare DecorRCFN and DecorRCGN by evaluating the differences in user satisfaction between them and CLIP-greedy, calculated by subtracting the user satisfaction levels of CLIP-greedy from those of DecorRCFN and DecorRCGN.
Figure \ref{fig:userStudy} displays the results of the experiments. 
At first round, these model exhibit similar performance due to no significant difference in their policies.
As the number of rounds increases, the difference in user satisfaction levels with the two models compared to CLIP-greedy increases in general, which indicates that DecorRCFN and DecorRCGN can better extract user preferences from history interaction.
Moreover, the performance of DecorRCFN outperforms DecorRCGN by an increasingly large margin.
The users explain that DecorRCGN tend to recommend the similar images as the previous round, which affects user satisfaction and gets lower rewards than DecorRCFN. 
They also comment that DecorRCGN has the best memory of user feedback, which could be explained by a fine-grained comparison with the history feedback and images.

\section{Conclusion}
In this paper, we proposed an Interactive Interior Design Recommendation System (IIDRS) based on reinforcement learning to mine user preferences in personalized interior design recommendation. We also proposed a Decoration Recommendation Coarse-to-Fine Policy Network (DecorRCFN) to enhance decision-making the agent's ability in large action spaces. Finally, we introduced an object-aware feedback generation model to augment the training process and optimize policy network training.
Our experiments demonstrated the superiority of our method in terms of recommendation accuracy and real-world user satisfaction. 
We anticipate this work will have a positive impact on personalized interior decoration design by introducing an interactive approach to recommendation.
We also acknowledge that current study may not meet users' requirements when their desired interior design cases do not exist in our dataset. In future work, we will explore developing generative decoration recommendation model that can create more personalized and feasible interior design cases based on  user preference.



\begin{acks}
This work was supported by the City-University Joint Funding Project of Guangzhou Science and Technology Plan (No. 2023A03J0141), Foshan HKUST Projects (FSUST21-FYTRI01A, FSUST21-FYTRI02A) and OPPO Research Fund.
\end{acks}

\bibliographystyle{ACM-Reference-Format}
\balance
\bibliography{sample-base}


\begin{thebibliography}{42}


\ifx \showCODEN    \undefined \def \showCODEN     #1{\unskip}     \fi
\ifx \showDOI      \undefined \def \showDOI       #1{#1}\fi
\ifx \showISBNx    \undefined \def \showISBNx     #1{\unskip}     \fi
\ifx \showISBNxiii \undefined \def \showISBNxiii  #1{\unskip}     \fi
\ifx \showISSN     \undefined \def \showISSN      #1{\unskip}     \fi
\ifx \showLCCN     \undefined \def \showLCCN      #1{\unskip}     \fi
\ifx \shownote     \undefined \def \shownote      #1{#1}          \fi
\ifx \showarticletitle \undefined \def \showarticletitle #1{#1}   \fi
\ifx \showURL      \undefined \def \showURL       {\relax}        \fi
\providecommand\bibfield[2]{#2}
\providecommand\bibinfo[2]{#2}
\providecommand\natexlab[1]{#1}
\providecommand\showeprint[2][]{arXiv:#2}

\bibitem[Chen et~al\mbox{.}(2019)]%
        {chen2019towards}
\bibfield{author}{\bibinfo{person}{Qibin Chen}, \bibinfo{person}{Junyang Lin},
  \bibinfo{person}{Yichang Zhang}, \bibinfo{person}{Ming Ding},
  \bibinfo{person}{Yukuo Cen}, \bibinfo{person}{Hongxia Yang}, {and}
  \bibinfo{person}{Jie Tang}.} \bibinfo{year}{2019}\natexlab{}.
\newblock \showarticletitle{Towards knowledge-based recommender dialog system}.
\newblock \bibinfo{journal}{\emph{arXiv preprint arXiv:1908.05391}}
  (\bibinfo{year}{2019}).
\newblock


\bibitem[Delmas et~al\mbox{.}(2022)]%
        {delmas2022artemis}
\bibfield{author}{\bibinfo{person}{Ginger Delmas},
  \bibinfo{person}{Rafael~Sampaio de Rezende}, \bibinfo{person}{Gabriela
  Csurka}, {and} \bibinfo{person}{Diane Larlus}.}
  \bibinfo{year}{2022}\natexlab{}.
\newblock \showarticletitle{ARTEMIS: Attention-based Retrieval with
  Text-Explicit Matching and Implicit Similarity}.
\newblock \bibinfo{journal}{\emph{arXiv preprint arXiv:2203.08101}}
  (\bibinfo{year}{2022}).
\newblock


\bibitem[Deng et~al\mbox{.}(2021b)]%
        {deng2021transvg}
\bibfield{author}{\bibinfo{person}{Jiajun Deng}, \bibinfo{person}{Zhengyuan
  Yang}, \bibinfo{person}{Tianlang Chen}, \bibinfo{person}{Wengang Zhou}, {and}
  \bibinfo{person}{Houqiang Li}.} \bibinfo{year}{2021}\natexlab{b}.
\newblock \showarticletitle{Transvg: End-to-end visual grounding with
  transformers}. In \bibinfo{booktitle}{\emph{Proceedings of the IEEE/CVF
  International Conference on Computer Vision}}. \bibinfo{pages}{1769--1779}.
\newblock


\bibitem[Deng et~al\mbox{.}(2021a)]%
        {deng2021unified}
\bibfield{author}{\bibinfo{person}{Yang Deng}, \bibinfo{person}{Yaliang Li},
  \bibinfo{person}{Fei Sun}, \bibinfo{person}{Bolin Ding}, {and}
  \bibinfo{person}{Wai Lam}.} \bibinfo{year}{2021}\natexlab{a}.
\newblock \showarticletitle{Unified conversational recommendation policy
  learning via graph-based reinforcement learning}. In
  \bibinfo{booktitle}{\emph{Proceedings of the 44th International ACM SIGIR
  Conference on Research and Development in Information Retrieval}}.
  \bibinfo{pages}{1431--1441}.
\newblock


\bibitem[Devlin et~al\mbox{.}(2018)]%
        {devlin2018bert}
\bibfield{author}{\bibinfo{person}{Jacob Devlin}, \bibinfo{person}{Ming-Wei
  Chang}, \bibinfo{person}{Kenton Lee}, {and} \bibinfo{person}{Kristina
  Toutanova}.} \bibinfo{year}{2018}\natexlab{}.
\newblock \showarticletitle{Bert: Pre-training of deep bidirectional
  transformers for language understanding}.
\newblock \bibinfo{journal}{\emph{arXiv preprint arXiv:1810.04805}}
  (\bibinfo{year}{2018}).
\newblock


\bibitem[Dhingra et~al\mbox{.}(2016)]%
        {DBLP:journals/corr/DhingraLLGCAD16}
\bibfield{author}{\bibinfo{person}{Bhuwan Dhingra}, \bibinfo{person}{Lihong
  Li}, \bibinfo{person}{Xiujun Li}, \bibinfo{person}{Jianfeng Gao},
  \bibinfo{person}{Yun{-}Nung Chen}, \bibinfo{person}{Faisal Ahmed}, {and}
  \bibinfo{person}{Li Deng}.} \bibinfo{year}{2016}\natexlab{}.
\newblock \showarticletitle{End-to-End Reinforcement Learning of Dialogue
  Agents for Information Access}.
\newblock \bibinfo{journal}{\emph{CoRR}}  \bibinfo{volume}{abs/1609.00777}
  (\bibinfo{year}{2016}).
\newblock
\showeprint[arXiv]{1609.00777}
\urldef\tempurl%
\url{http://arxiv.org/abs/1609.00777}
\showURL{%
\tempurl}


\bibitem[Di and Yu(2021)]%
        {di2021multi}
\bibfield{author}{\bibinfo{person}{Xinhan Di} {and} \bibinfo{person}{Pengqian
  Yu}.} \bibinfo{year}{2021}\natexlab{}.
\newblock \showarticletitle{Multi-Agent Reinforcement Learning of 3D Furniture
  Layout Simulation in Indoor Graphics Scenes}.
\newblock \bibinfo{journal}{\emph{arXiv preprint arXiv:2102.09137}}
  (\bibinfo{year}{2021}).
\newblock


\bibitem[Gu et~al\mbox{.}(2022)]%
        {gu2022vector}
\bibfield{author}{\bibinfo{person}{Shuyang Gu}, \bibinfo{person}{Dong Chen},
  \bibinfo{person}{Jianmin Bao}, \bibinfo{person}{Fang Wen},
  \bibinfo{person}{Bo Zhang}, \bibinfo{person}{Dongdong Chen},
  \bibinfo{person}{Lu Yuan}, {and} \bibinfo{person}{Baining Guo}.}
  \bibinfo{year}{2022}\natexlab{}.
\newblock \showarticletitle{Vector quantized diffusion model for text-to-image
  synthesis}. In \bibinfo{booktitle}{\emph{Proceedings of the IEEE/CVF
  Conference on Computer Vision and Pattern Recognition}}.
  \bibinfo{pages}{10696--10706}.
\newblock


\bibitem[Guo et~al\mbox{.}(2022)]%
        {guo2022context}
\bibfield{author}{\bibinfo{person}{Weiyu Guo}, \bibinfo{person}{Zhaoshuo Li},
  \bibinfo{person}{Yongkui Yang}, \bibinfo{person}{Zheng Wang},
  \bibinfo{person}{Russell~H Taylor}, \bibinfo{person}{Mathias Unberath},
  \bibinfo{person}{Alan Yuille}, {and} \bibinfo{person}{Yingwei Li}.}
  \bibinfo{year}{2022}\natexlab{}.
\newblock \showarticletitle{Context-Enhanced Stereo Transformer}. In
  \bibinfo{booktitle}{\emph{Computer Vision--ECCV 2022: 17th European
  Conference, Tel Aviv, Israel, October 23--27, 2022, Proceedings, Part
  XXXII}}. Springer, \bibinfo{pages}{263--279}.
\newblock


\bibitem[Guo et~al\mbox{.}(2018)]%
        {guo2018dialog}
\bibfield{author}{\bibinfo{person}{Xiaoxiao Guo}, \bibinfo{person}{Hui Wu},
  \bibinfo{person}{Yu Cheng}, \bibinfo{person}{Steven Rennie},
  \bibinfo{person}{Gerald Tesauro}, {and} \bibinfo{person}{Rogerio Feris}.}
  \bibinfo{year}{2018}\natexlab{}.
\newblock \showarticletitle{Dialog-based interactive image retrieval}.
\newblock \bibinfo{journal}{\emph{Advances in neural information processing
  systems}}  \bibinfo{volume}{31} (\bibinfo{year}{2018}).
\newblock


\bibitem[H{\"o}llein et~al\mbox{.}(2023)]%
        {hollein2023text2room}
\bibfield{author}{\bibinfo{person}{Lukas H{\"o}llein}, \bibinfo{person}{Ang
  Cao}, \bibinfo{person}{Andrew Owens}, \bibinfo{person}{Justin Johnson}, {and}
  \bibinfo{person}{Matthias Nie{\ss}ner}.} \bibinfo{year}{2023}\natexlab{}.
\newblock \showarticletitle{Text2Room: Extracting Textured 3D Meshes from 2D
  Text-to-Image Models}.
\newblock \bibinfo{journal}{\emph{arXiv preprint arXiv:2303.11989}}
  (\bibinfo{year}{2023}).
\newblock


\bibitem[Hossain et~al\mbox{.}(2019)]%
        {hossain2019comprehensive}
\bibfield{author}{\bibinfo{person}{MD~Zakir Hossain}, \bibinfo{person}{Ferdous
  Sohel}, \bibinfo{person}{Mohd~Fairuz Shiratuddin}, {and}
  \bibinfo{person}{Hamid Laga}.} \bibinfo{year}{2019}\natexlab{}.
\newblock \showarticletitle{A comprehensive survey of deep learning for image
  captioning}.
\newblock \bibinfo{journal}{\emph{ACM Computing Surveys (CsUR)}}
  \bibinfo{volume}{51}, \bibinfo{number}{6} (\bibinfo{year}{2019}),
  \bibinfo{pages}{1--36}.
\newblock


\bibitem[Jawaheer et~al\mbox{.}(2014)]%
        {jawaheer2014modeling}
\bibfield{author}{\bibinfo{person}{Gawesh Jawaheer}, \bibinfo{person}{Peter
  Weller}, {and} \bibinfo{person}{Patty Kostkova}.}
  \bibinfo{year}{2014}\natexlab{}.
\newblock \showarticletitle{Modeling user preferences in recommender systems: A
  classification framework for explicit and implicit user feedback}.
\newblock \bibinfo{journal}{\emph{ACM Transactions on Interactive Intelligent
  Systems (TiiS)}} \bibinfo{volume}{4}, \bibinfo{number}{2}
  (\bibinfo{year}{2014}), \bibinfo{pages}{1--26}.
\newblock


\bibitem[Kotkov et~al\mbox{.}(2018)]%
        {kotkov2018investigating}
\bibfield{author}{\bibinfo{person}{Denis Kotkov}, \bibinfo{person}{Joseph~A
  Konstan}, \bibinfo{person}{Qian Zhao}, {and} \bibinfo{person}{Jari
  Veijalainen}.} \bibinfo{year}{2018}\natexlab{}.
\newblock \showarticletitle{Investigating serendipity in recommender systems
  based on real user feedback}. In \bibinfo{booktitle}{\emph{Proceedings of the
  33rd annual acm symposium on applied computing}}.
  \bibinfo{pages}{1341--1350}.
\newblock


\bibitem[Lei et~al\mbox{.}(2020a)]%
        {lei2020estimation}
\bibfield{author}{\bibinfo{person}{Wenqiang Lei}, \bibinfo{person}{Xiangnan
  He}, \bibinfo{person}{Yisong Miao}, \bibinfo{person}{Qingyun Wu},
  \bibinfo{person}{Richang Hong}, \bibinfo{person}{Min-Yen Kan}, {and}
  \bibinfo{person}{Tat-Seng Chua}.} \bibinfo{year}{2020}\natexlab{a}.
\newblock \showarticletitle{Estimation-action-reflection: Towards deep
  interaction between conversational and recommender systems}. In
  \bibinfo{booktitle}{\emph{Proceedings of the 13th International Conference on
  Web Search and Data Mining}}. \bibinfo{pages}{304--312}.
\newblock


\bibitem[Lei et~al\mbox{.}(2020b)]%
        {lei2020interactive}
\bibfield{author}{\bibinfo{person}{Wenqiang Lei}, \bibinfo{person}{Gangyi
  Zhang}, \bibinfo{person}{Xiangnan He}, \bibinfo{person}{Yisong Miao},
  \bibinfo{person}{Xiang Wang}, \bibinfo{person}{Liang Chen}, {and}
  \bibinfo{person}{Tat-Seng Chua}.} \bibinfo{year}{2020}\natexlab{b}.
\newblock \showarticletitle{Interactive path reasoning on graph for
  conversational recommendation}. In \bibinfo{booktitle}{\emph{Proceedings of
  the 26th ACM SIGKDD international conference on knowledge discovery \& data
  mining}}. \bibinfo{pages}{2073--2083}.
\newblock


\bibitem[Li et~al\mbox{.}(2018)]%
        {li2018towards}
\bibfield{author}{\bibinfo{person}{Raymond Li}, \bibinfo{person}{Samira
  Ebrahimi~Kahou}, \bibinfo{person}{Hannes Schulz}, \bibinfo{person}{Vincent
  Michalski}, \bibinfo{person}{Laurent Charlin}, {and} \bibinfo{person}{Chris
  Pal}.} \bibinfo{year}{2018}\natexlab{}.
\newblock \showarticletitle{Towards deep conversational recommendations}.
\newblock \bibinfo{journal}{\emph{Advances in neural information processing
  systems}}  \bibinfo{volume}{31} (\bibinfo{year}{2018}).
\newblock


\bibitem[Liao et~al\mbox{.}(2018)]%
        {liao2018interpretable}
\bibfield{author}{\bibinfo{person}{Lizi Liao}, \bibinfo{person}{Xiangnan He},
  \bibinfo{person}{Bo Zhao}, \bibinfo{person}{Chong-Wah Ngo}, {and}
  \bibinfo{person}{Tat-Seng Chua}.} \bibinfo{year}{2018}\natexlab{}.
\newblock \showarticletitle{Interpretable multimodal retrieval for fashion
  products}. In \bibinfo{booktitle}{\emph{Proceedings of the 26th ACM
  international conference on Multimedia}}. \bibinfo{pages}{1571--1579}.
\newblock


\bibitem[Long et~al\mbox{.}(2016)]%
        {long2016composite}
\bibfield{author}{\bibinfo{person}{Mingsheng Long}, \bibinfo{person}{Yue Cao},
  \bibinfo{person}{Jianmin Wang}, {and} \bibinfo{person}{Philip~S Yu}.}
  \bibinfo{year}{2016}\natexlab{}.
\newblock \showarticletitle{Composite correlation quantization for efficient
  multimodal retrieval}. In \bibinfo{booktitle}{\emph{Proceedings of the 39th
  International ACM SIGIR conference on Research and Development in Information
  Retrieval}}. \bibinfo{pages}{579--588}.
\newblock


\bibitem[Mildenhall et~al\mbox{.}(2021)]%
        {mildenhall2021nerf}
\bibfield{author}{\bibinfo{person}{Ben Mildenhall}, \bibinfo{person}{Pratul~P
  Srinivasan}, \bibinfo{person}{Matthew Tancik}, \bibinfo{person}{Jonathan~T
  Barron}, \bibinfo{person}{Ravi Ramamoorthi}, {and} \bibinfo{person}{Ren Ng}.}
  \bibinfo{year}{2021}\natexlab{}.
\newblock \showarticletitle{Nerf: Representing scenes as neural radiance fields
  for view synthesis}.
\newblock \bibinfo{journal}{\emph{Commun. ACM}} \bibinfo{volume}{65},
  \bibinfo{number}{1} (\bibinfo{year}{2021}), \bibinfo{pages}{99--106}.
\newblock


\bibitem[Radford et~al\mbox{.}(2021)]%
        {radford2021learning}
\bibfield{author}{\bibinfo{person}{Alec Radford}, \bibinfo{person}{Jong~Wook
  Kim}, \bibinfo{person}{Chris Hallacy}, \bibinfo{person}{Aditya Ramesh},
  \bibinfo{person}{Gabriel Goh}, \bibinfo{person}{Sandhini Agarwal},
  \bibinfo{person}{Girish Sastry}, \bibinfo{person}{Amanda Askell},
  \bibinfo{person}{Pamela Mishkin}, \bibinfo{person}{Jack Clark},
  {et~al\mbox{.}}} \bibinfo{year}{2021}\natexlab{}.
\newblock \showarticletitle{Learning transferable visual models from natural
  language supervision}. In \bibinfo{booktitle}{\emph{International Conference
  on Machine Learning}}. PMLR, \bibinfo{pages}{8748--8763}.
\newblock


\bibitem[Ramesh et~al\mbox{.}(2021)]%
        {ramesh2021zero}
\bibfield{author}{\bibinfo{person}{Aditya Ramesh}, \bibinfo{person}{Mikhail
  Pavlov}, \bibinfo{person}{Gabriel Goh}, \bibinfo{person}{Scott Gray},
  \bibinfo{person}{Chelsea Voss}, \bibinfo{person}{Alec Radford},
  \bibinfo{person}{Mark Chen}, {and} \bibinfo{person}{Ilya Sutskever}.}
  \bibinfo{year}{2021}\natexlab{}.
\newblock \showarticletitle{Zero-shot text-to-image generation}. In
  \bibinfo{booktitle}{\emph{International Conference on Machine Learning}}.
  PMLR, \bibinfo{pages}{8821--8831}.
\newblock


\bibitem[Sharma et~al\mbox{.}(2018)]%
        {sharma2018chatpainter}
\bibfield{author}{\bibinfo{person}{Shikhar Sharma}, \bibinfo{person}{Dendi
  Suhubdy}, \bibinfo{person}{Vincent Michalski},
  \bibinfo{person}{Samira~Ebrahimi Kahou}, {and} \bibinfo{person}{Yoshua
  Bengio}.} \bibinfo{year}{2018}\natexlab{}.
\newblock \showarticletitle{Chatpainter: Improving text to image generation
  using dialogue}.
\newblock \bibinfo{journal}{\emph{arXiv preprint arXiv:1802.08216}}
  (\bibinfo{year}{2018}).
\newblock


\bibitem[Stefanini et~al\mbox{.}(2022)]%
        {stefanini2022show}
\bibfield{author}{\bibinfo{person}{Matteo Stefanini}, \bibinfo{person}{Marcella
  Cornia}, \bibinfo{person}{Lorenzo Baraldi}, \bibinfo{person}{Silvia
  Cascianelli}, \bibinfo{person}{Giuseppe Fiameni}, {and} \bibinfo{person}{Rita
  Cucchiara}.} \bibinfo{year}{2022}\natexlab{}.
\newblock \showarticletitle{From show to tell: a survey on deep learning-based
  image captioning}.
\newblock \bibinfo{journal}{\emph{IEEE transactions on pattern analysis and
  machine intelligence}} \bibinfo{volume}{45}, \bibinfo{number}{1}
  (\bibinfo{year}{2022}), \bibinfo{pages}{539--559}.
\newblock


\bibitem[Sun and Zhang(2018)]%
        {sun2018conversational}
\bibfield{author}{\bibinfo{person}{Yueming Sun} {and} \bibinfo{person}{Yi
  Zhang}.} \bibinfo{year}{2018}\natexlab{}.
\newblock \showarticletitle{Conversational recommender system}. In
  \bibinfo{booktitle}{\emph{The 41st international acm sigir conference on
  research \& development in information retrieval}}.
  \bibinfo{pages}{235--244}.
\newblock


\bibitem[Tan et~al\mbox{.}(2019)]%
        {tan2019text2scene}
\bibfield{author}{\bibinfo{person}{Fuwen Tan}, \bibinfo{person}{Song Feng},
  {and} \bibinfo{person}{Vicente Ordonez}.} \bibinfo{year}{2019}\natexlab{}.
\newblock \showarticletitle{Text2scene: Generating compositional scenes from
  textual descriptions}. In \bibinfo{booktitle}{\emph{Proceedings of the
  IEEE/CVF Conference on Computer Vision and Pattern Recognition}}.
  \bibinfo{pages}{6710--6719}.
\newblock


\bibitem[Tan and Bansal(2019)]%
        {tan2019lxmert}
\bibfield{author}{\bibinfo{person}{Hao Tan} {and} \bibinfo{person}{Mohit
  Bansal}.} \bibinfo{year}{2019}\natexlab{}.
\newblock \showarticletitle{LXMERT: Learning Cross-Modality Encoder
  Representations from Transformers}. In \bibinfo{booktitle}{\emph{Proceedings
  of the 2019 Conference on Empirical Methods in Natural Language Processing}}.
\newblock


\bibitem[Tang et~al\mbox{.}(2023)]%
        {tang2023diffuscene}
\bibfield{author}{\bibinfo{person}{Jiapeng Tang}, \bibinfo{person}{Yinyu Nie},
  \bibinfo{person}{Lev Markhasin}, \bibinfo{person}{Angela Dai},
  \bibinfo{person}{Justus Thies}, {and} \bibinfo{person}{Matthias
  Nie{\ss}ner}.} \bibinfo{year}{2023}\natexlab{}.
\newblock \showarticletitle{DiffuScene: Scene Graph Denoising Diffusion
  Probabilistic Model for Generative Indoor Scene Synthesis}.
\newblock \bibinfo{journal}{\emph{arXiv preprint arXiv:2303.14207}}
  (\bibinfo{year}{2023}).
\newblock


\bibitem[Turk(2014)]%
        {turk2014multimodal}
\bibfield{author}{\bibinfo{person}{Matthew Turk}.}
  \bibinfo{year}{2014}\natexlab{}.
\newblock \showarticletitle{Multimodal interaction: A review}.
\newblock \bibinfo{journal}{\emph{Pattern recognition letters}}
  \bibinfo{volume}{36} (\bibinfo{year}{2014}), \bibinfo{pages}{189--195}.
\newblock


\bibitem[Vaswani et~al\mbox{.}(2017)]%
        {vaswani2017attention}
\bibfield{author}{\bibinfo{person}{Ashish Vaswani}, \bibinfo{person}{Noam
  Shazeer}, \bibinfo{person}{Niki Parmar}, \bibinfo{person}{Jakob Uszkoreit},
  \bibinfo{person}{Llion Jones}, \bibinfo{person}{Aidan~N Gomez},
  \bibinfo{person}{{\L}ukasz Kaiser}, {and} \bibinfo{person}{Illia
  Polosukhin}.} \bibinfo{year}{2017}\natexlab{}.
\newblock \showarticletitle{Attention is all you need}.
\newblock \bibinfo{journal}{\emph{Advances in neural information processing
  systems}}  \bibinfo{volume}{30} (\bibinfo{year}{2017}).
\newblock


\bibitem[Wang et~al\mbox{.}(2022a)]%
        {wang2022neuris}
\bibfield{author}{\bibinfo{person}{Jiepeng Wang}, \bibinfo{person}{Peng Wang},
  \bibinfo{person}{Xiaoxiao Long}, \bibinfo{person}{Christian Theobalt},
  \bibinfo{person}{Taku Komura}, \bibinfo{person}{Lingjie Liu}, {and}
  \bibinfo{person}{Wenping Wang}.} \bibinfo{year}{2022}\natexlab{a}.
\newblock \showarticletitle{Neuris: Neural reconstruction of indoor scenes
  using normal priors}. In \bibinfo{booktitle}{\emph{Computer Vision--ECCV
  2022: 17th European Conference, Tel Aviv, Israel, October 23--27, 2022,
  Proceedings, Part XXXII}}. Springer, \bibinfo{pages}{139--155}.
\newblock


\bibitem[Wang et~al\mbox{.}(2022b)]%
        {wang2022ofa}
\bibfield{author}{\bibinfo{person}{Peng Wang}, \bibinfo{person}{An Yang},
  \bibinfo{person}{Rui Men}, \bibinfo{person}{Junyang Lin},
  \bibinfo{person}{Shuai Bai}, \bibinfo{person}{Zhikang Li},
  \bibinfo{person}{Jianxin Ma}, \bibinfo{person}{Chang Zhou},
  \bibinfo{person}{Jingren Zhou}, {and} \bibinfo{person}{Hongxia Yang}.}
  \bibinfo{year}{2022}\natexlab{b}.
\newblock \showarticletitle{Ofa: Unifying architectures, tasks, and modalities
  through a simple sequence-to-sequence learning framework}. In
  \bibinfo{booktitle}{\emph{International Conference on Machine Learning}}.
  PMLR, \bibinfo{pages}{23318--23340}.
\newblock


\bibitem[Wang et~al\mbox{.}(2021)]%
        {wang2021sceneformer}
\bibfield{author}{\bibinfo{person}{Xinpeng Wang}, \bibinfo{person}{Chandan
  Yeshwanth}, {and} \bibinfo{person}{Matthias Nie{\ss}ner}.}
  \bibinfo{year}{2021}\natexlab{}.
\newblock \showarticletitle{Sceneformer: Indoor scene generation with
  transformers}. In \bibinfo{booktitle}{\emph{2021 International Conference on
  3D Vision (3DV)}}. IEEE, \bibinfo{pages}{106--115}.
\newblock


\bibitem[Wei et~al\mbox{.}(2023)]%
        {wei2023lego}
\bibfield{author}{\bibinfo{person}{Qiuhong~Anna Wei}, \bibinfo{person}{Sijie
  Ding}, \bibinfo{person}{Jeong~Joon Park}, \bibinfo{person}{Rahul Sajnani},
  \bibinfo{person}{Adrien Poulenard}, \bibinfo{person}{Srinath Sridhar}, {and}
  \bibinfo{person}{Leonidas Guibas}.} \bibinfo{year}{2023}\natexlab{}.
\newblock \showarticletitle{Lego-net: Learning regular rearrangements of
  objects in rooms}.
\newblock \bibinfo{journal}{\emph{arXiv preprint arXiv:2301.09629}}
  (\bibinfo{year}{2023}).
\newblock


\bibitem[Wei et~al\mbox{.}(2020)]%
        {wei2020multi}
\bibfield{author}{\bibinfo{person}{Xi Wei}, \bibinfo{person}{Tianzhu Zhang},
  \bibinfo{person}{Yan Li}, \bibinfo{person}{Yongdong Zhang}, {and}
  \bibinfo{person}{Feng Wu}.} \bibinfo{year}{2020}\natexlab{}.
\newblock \showarticletitle{Multi-modality cross attention network for image
  and sentence matching}. In \bibinfo{booktitle}{\emph{Proceedings of the
  IEEE/CVF conference on computer vision and pattern recognition}}.
  \bibinfo{pages}{10941--10950}.
\newblock


\bibitem[Williams(1992)]%
        {williams1992simple}
\bibfield{author}{\bibinfo{person}{Ronald~J Williams}.}
  \bibinfo{year}{1992}\natexlab{}.
\newblock \showarticletitle{Simple statistical gradient-following algorithms
  for connectionist reinforcement learning}.
\newblock \bibinfo{journal}{\emph{Reinforcement learning}}
  (\bibinfo{year}{1992}), \bibinfo{pages}{5--32}.
\newblock


\bibitem[Xu et~al\mbox{.}(2021)]%
        {xu2021adapting}
\bibfield{author}{\bibinfo{person}{Kerui Xu}, \bibinfo{person}{Jingxuan Yang},
  \bibinfo{person}{Jun Xu}, \bibinfo{person}{Sheng Gao}, \bibinfo{person}{Jun
  Guo}, {and} \bibinfo{person}{Ji-Rong Wen}.} \bibinfo{year}{2021}\natexlab{}.
\newblock \showarticletitle{Adapting user preference to online feedback in
  multi-round conversational recommendation}. In
  \bibinfo{booktitle}{\emph{Proceedings of the 14th ACM international
  conference on web search and data mining}}. \bibinfo{pages}{364--372}.
\newblock


\bibitem[Yang et~al\mbox{.}(2022)]%
        {yang2022chinese}
\bibfield{author}{\bibinfo{person}{An Yang}, \bibinfo{person}{Junshu Pan},
  \bibinfo{person}{Junyang Lin}, \bibinfo{person}{Rui Men},
  \bibinfo{person}{Yichang Zhang}, \bibinfo{person}{Jingren Zhou}, {and}
  \bibinfo{person}{Chang Zhou}.} \bibinfo{year}{2022}\natexlab{}.
\newblock \showarticletitle{Chinese CLIP: Contrastive Vision-Language
  Pretraining in Chinese}.
\newblock \bibinfo{journal}{\emph{arXiv preprint arXiv:2211.01335}}
  (\bibinfo{year}{2022}).
\newblock


\bibitem[Yuan and Lam(2021)]%
        {yuan2021conversational}
\bibfield{author}{\bibinfo{person}{Yifei Yuan} {and} \bibinfo{person}{Wai
  Lam}.} \bibinfo{year}{2021}\natexlab{}.
\newblock \showarticletitle{Conversational fashion image retrieval via
  multiturn natural language feedback}. In
  \bibinfo{booktitle}{\emph{Proceedings of the 44th International ACM SIGIR
  Conference on Research and Development in Information Retrieval}}.
  \bibinfo{pages}{839--848}.
\newblock


\bibitem[Zhang et~al\mbox{.}(2020)]%
        {zhang2020reward}
\bibfield{author}{\bibinfo{person}{Ruiyi Zhang}, \bibinfo{person}{Tong Yu},
  \bibinfo{person}{Yilin Shen}, \bibinfo{person}{Hongxia Jin},
  \bibinfo{person}{Changyou Chen}, {and} \bibinfo{person}{Lawrence Carin}.}
  \bibinfo{year}{2020}\natexlab{}.
\newblock \showarticletitle{Reward Constrained Interactive Recommendation with
  Natural Language Feedback}.
\newblock \bibinfo{journal}{\emph{arXiv preprint arXiv:2005.01618}}
  (\bibinfo{year}{2020}).
\newblock


\bibitem[Zhou et~al\mbox{.}(2020a)]%
        {zhou2020improving}
\bibfield{author}{\bibinfo{person}{Kun Zhou}, \bibinfo{person}{Wayne~Xin Zhao},
  \bibinfo{person}{Shuqing Bian}, \bibinfo{person}{Yuanhang Zhou},
  \bibinfo{person}{Ji-Rong Wen}, {and} \bibinfo{person}{Jingsong Yu}.}
  \bibinfo{year}{2020}\natexlab{a}.
\newblock \showarticletitle{Improving conversational recommender systems via
  knowledge graph based semantic fusion}. In
  \bibinfo{booktitle}{\emph{Proceedings of the 26th ACM SIGKDD International
  Conference on Knowledge Discovery \& Data Mining}}.
  \bibinfo{pages}{1006--1014}.
\newblock


\bibitem[Zhou et~al\mbox{.}(2020b)]%
        {zhou2020towards}
\bibfield{author}{\bibinfo{person}{Kun Zhou}, \bibinfo{person}{Yuanhang Zhou},
  \bibinfo{person}{Wayne~Xin Zhao}, \bibinfo{person}{Xiaoke Wang}, {and}
  \bibinfo{person}{Ji-Rong Wen}.} \bibinfo{year}{2020}\natexlab{b}.
\newblock \showarticletitle{Towards topic-guided conversational recommender
  system}.
\newblock \bibinfo{journal}{\emph{arXiv preprint arXiv:2010.04125}}
  (\bibinfo{year}{2020}).
\newblock


\end{thebibliography}







\end{document}